**Vague Knowledge: Information without Transitivity and Partitions**

**Kerry Xiao**[1]

Hong Kong University of Science and Technology

ackerry@ust.hk

January 2026

**Abstract**

I relax the standard assumptions of transitivity and partition structure in economic models of information to formalize vague knowledge—non-transitive indistinguishability over states. I show that vague knowledge, while failing to partition the state space, remains informative by distinguishing some states from others. Moreover, it can only be faithfully expressed through vague communication with blurred boundaries. My results provide microfoundations for the prevalence of natural language communication and qualitative reasoning in the real world, where knowledge is often vague.

---

[1] Department of Accounting. The Hong Kong University of Science and Technology Clear Water Bay, Kowloon, Hong Kong.

The author has no relevant financial or non-financial interests to disclose.

**Vague Knowledge: Information without Transitivity and Partitions**

## 1. Introduction

Standard economic theory models information as a partition of a state space (Osborne and Rubinstein 1994; Samuelson 2004; Bikhchandani, Hirshleifer and Riley 2013). In this framework, all information is precise: each signal cleanly separates certain states from all others, implying that the indistinguishability relation over states is transitive. Any signal failing to partition the state space is deemed uninformative. However, much of real-world knowledge is vague and not quantitatively measurable (Keynes 1921; Knight 1921; Hayek 1974). Is such vague knowledge truly uninformative? The rise of large language models demonstrates the value of vague, qualitative information. In this study, I relax the assumptions of transitivity and partition structures to show how knowledge can be vague yet informative.

I model an individual's uncertainty as a binary indistinguishability relation over states of the world (Figure 1A). Observing a difference between two states eliminates their indistinguishability, thereby generates knowledge. Knowledge is precise if the indistinguishability relation is transitive; otherwise it is vague. Transitivity requires that if a state can be distinguished from one member of a group, it can be distinguished from all members. Precise knowledge allows the individual to pinpoint specific states relative to all others (Figure 1B). In contrast, vague knowledge occurs when the individual can *distinguish certain states from some others but not from all*.

[Insert Figure 1 Here]

Human perceptual limits (and machine detection limits) often create situations where large differences between states can be discerned, but small differences cannot.[1] For example, an individual might distinguish state $a$ from a clearly different state $c$, yet fail to distinguish $a$ from a similar state $b$, and $b$ from $c$ (Figure 1C). Here, distinguishing $a$ from $c$ does not extend to $a$ versus $b$, even though $b$ and $c$ are indistinguishable. Thus, the elimination of indistinguishability is non-transitive. The resulting knowledge is vague, failing to partition the state space. Yet, it is informative, separating previously indistinguishable states ($a$ versus $c$) and thereby reducing uncertainty.

The precision of knowledge constrains how precisely it can be expressed. An expression is precise if it corresponds to an information set with a sharp boundary; otherwise, the expression is vague. In my model, precise knowledge induces a partitional information structure: transitively indistinguishable states form disjoint equivalence classes with well-defined boundaries, enabling precise expressions. By contrast, vague knowledge induces an information structure that is a cover of the state space rather than a partition: indistinguishable states cluster into groups that can overlap at the margins due to non-transitivity. Each such group has a core of states that are mutually indistinguishable, but also includes borderline cases that are partially indistinguishable from the core and from other groups.

Any expression that faithfully represents vague knowledge must therefore have a blurred boundary: it includes all core states and some borderline states. In other words, the expression comes with *a boundary region to admit borderline cases*. Vague knowledge can thus only be communicated vaguely. Indeed, in the real world, natural language provides this kind of boundary-blurring communication: every word has a typical meaning for core states but can extend to borderline states depending on context.[2] This inherent vagueness of natural language allows individuals to convey knowledge that cannot be pinned down with numerical precision.

[1] For epistemic vagueness, see Williamson (1994) and van Deemter (2010).

**Related Literature.** Despite being widely studied in other fields, vagueness remains underexplored in economics.[3] Lipman (2025) argues that individuals often hold vague views rather than merely expressing precise views vaguely, and thus calls for a new framework for understanding such information and knowledge. In response, I explicitly model vague knowledge and show that it naturally leads to vague communication. I introduce precision (lack of vagueness) as a new dimension of informativeness: mutually exclusive information sets indicate precise information, whereas overlapping sets indicate vague information. This departs from the classical partition-based measure of informativeness, where more, smaller information sets mean finer information (Samuelson 2004). Methodologically, I represent information as a binary relation over states (Rubinstein 1996), without assuming transitivity or a partitional structure. Non-transitive relations are well-known in preference theory (e.g., Luce 1956) but are new to knowledge modeling in economics.[4] Furthermore, unlike models of incomplete knowledge (e.g., Samet 1990; Dekel, Lipman, and Rustichini 1998; Geanakoplos 2021), I show that even with complete awareness of all states, imprecise knowledge can still yield a non-partitional information structure.[5]

My model offers microfoundations for the prevalence of natural language and soft information in the real world. Non-transitive indistinguishability captures scenarios where individuals observe some attributes of an event but not enough to define it precisely. This leads to non-partitional information structures that impede quantitative measurement and probabilistic reasoning (Keynes 1921; Knight 1921; Hayek 1974) but still permit qualitative reasoning and communication (Wallsten 1990; Kay and King 2020). Because real-world knowledge is often imprecise, qualitative and vague approaches are common: for example, central banks and firms issue qualitative guidance when uncertainty is high; R&D evaluations and venture capital contracts rely on subjective judgment; and regulators prefer principles over rules in emerging domains like AI and climate change. Classical models that assume transitivity and partitioned information struggle to account for these practices. In contrast, my model suggests that ignoring vague knowledge and insisting on an overly precise, quantitative approach can be counterproductive in such settings.

[2] For linguistic vagueness, see Russell (1923), Wittgenstein (1953), and Solt (2015).

[3] See reviews by Keefe (2000) in philosophy; Solt (2015) in linguistics; van Deemter (2010) in computational linguistics; Pinker (1997) in psychology; Wallsten (1990) and Kay and King (2020) in decision science; and Endicott (2000) in legal research.

[4] In epistemic logic, vague knowledge is often modeled with a non-transitive accessibility relation (Williamson 1994). I adapt this insight to economics, showing that a non-transitive indistinguishability relation over states produces a non-partitional information structure.

[5] Models of incomplete knowledge relax the partition assumption via possibility correspondences. In contrast, I model vague knowledge with a non-transitive indistinguishability relation. Thus, knowledge in my framework is not absent but imprecise, yet it remains informative and can be communicated through natural language.

On the other hand, my model highlights a limitation of vague communication: a large boundary region (many borderline cases) makes interpretation highly context-dependent (Liberti and Petersen 2019). Thus, vague information can be interpreted inconsistently across agents, creating information asymmetries even when it is publicly available. This suggests that no single individual can fully aggregate vague information, which is why markets often serve as the most effective aggregators (Hayek 1945).

## 2. The Model

### 2.1 Vague Knowledge: Non-Transitive Relations

Let $\Omega$ be the set of all possible states of the world, where each state $\omega \in \Omega$ is a complete description of fundamentals. An individual's information is represented by a binary indistinguishability relation "$\sim$", which specifies which states she lacks the ability to discern any differences between. Formally:

**Definition 1 (Precise versus Vague Knowledge).** *An indistinguishability relation* $\sim$ *on* $\Omega$ *is a binary relation that is reflexive and symmetric, but not necessarily transitive. Let*

$$U = \{(\omega_1, \omega_2) \in \Omega \times \Omega | \omega_1 \sim \omega_2\}$$

*be the set of all indistinguishable state pairs. The individual's knowledge is then the set of all distinguishable state pairs,* $\Omega \times \Omega \backslash U$. *I classify knowledge as:*

- *Precise knowledge, if* $\sim$ *is transitive (i.e., an equivalence relation).*
- *Vague knowledge, if* $\sim$ *is non-transitive (i.e., a similarity relation).*

In a baseline scenario where the individual lacks knowledge, every pair of states is indistinguishable. As she acquires information through observation, she notices differences between states. Each observed difference eliminates the indistinguishability of that pair of states, shrinking $U$ to a proper subset of $\Omega \times \Omega$. These eliminations induce a grouping of states: states that remain mutually indistinguishable cluster together.

If the indistinguishability relation is transitive, the clusters of indistinguishable states are crisp. In this case, eliminating indistinguishabilities partitions $\Omega$ into disjoint equivalence classes. Each state is either completely indistinguishable from an entire class or completely distinguishable from that class. The individual knows exactly which states belong in each group. This is the classical case in economics.

Transitivity, however, is a strong condition that often fails in reality. Continuing the earlier example, suppose adjacent states are indistinguishable ($a \sim b$ and $b \sim c$) while non-adjacent states are distinguishable ($a \nsim c$). The individual cannot form an equivalence class around $a$. Instead, $a$ belongs to a similarity class (or an approximation class) that blurs into others via the borderline state $b$. In general, an individual with limited discriminatory ability will have vague knowledge. The induced information structure is a cover of $\Omega$ rather than a partition: groups of states can overlap through borderline cases.

## 2.2 Vague Expression: Non-Partitional Information Structures

Next, I formalize how knowledge is expressed. An information structure can consist of multiple information sets. For clarity, consider a single information set that the individual might describe. To allow for vagueness, I characterize this information set by its lower and upper boundaries, in the spirit of rough set theory (Pawlak 1982).

**Definition 2 (Precise versus Vague Expression)**. *An information set is a non-empty set of states $X \subseteq \Omega$ such that $\underline{X} \subseteq X \subseteq \overline{X}$. I define the lower boundary $\underline{X}$ and the upper boundary $\overline{X}$ as follows:*

*$\underline{X}$ is a subset $X$ of chosen to satisfy two conditions:*

*(1) Mutual indistinguishability: $\forall \omega_1, \omega_2 \in \underline{X}, \omega_1 \sim \omega_2$; and*

*(2) Relative maximality: $\forall \omega' \in X \backslash \underline{X}$, $\exists\ \omega \in \underline{X}$ such that $\omega \nsim \omega'$.*

*For a chosen $\underline{X}$,*

$$\overline{X} = \{\omega_1 \in \Omega \big| \exists \omega_2 \in \underline{X}, \omega_1 \sim \omega_2\}.$$

*I classify $X$ as:*

- *Precise information set, if $\underline{X} = \overline{X}$ (i.e. $X$ has a sharp boundary), and*
- *Vague information set, if $\underline{X} \subset \overline{X}$ (i.e. $X$ has a non-empty boundary region).*

The lower boundary $\underline{X}$ consists of the core states of $X$ – states that are mutually indistinguishable (i.e. they share a common feature).[6] In some cases, $\underline{X}$ may be a singleton. The upper boundary $\overline{X}$ consists of all states that are indistinguishable from at least one core state in $\underline{X}$.[7] This includes borderline states, both those within $X$ (indistinguishable from core states but not themselves core) and those outside $X$. Intuitively, $X$ *certainly* describes any states in $\underline{X}$, and *possibly* describes any state in $\overline{X}$ (since such a state cannot be ruled out given the information $X$ conveys). Equivalently, $X$ definitely does not describe any state outside $\overline{X}$.

[6] For a given relation $\sim$ and information set $X$, there may be multiple choices of $\underline{X}$. In such case, vagueness coexists with ambiguity (Ellsberg 1961). To focus on vagueness, I fix an arbitrary choice of $\underline{X}$ and define $\overline{X}$ relative to that choice. None of the subsequent results depends on which particular $\underline{X}$ is selected.

[7] For simplicity, I consider a single-layer outward extension of indistinguishability.

If the lower and upper boundaries coincide ($\underline{X} = X = \overline{X}$), then $X$ has a well-defined, sharp boundary, forming a partitional cell of $\Omega$. Quantitative expressions work this way, referring to sets of states with exact definitions. This is the standard assumption in economics. By contrast, if $\underline{X} \subset \overline{X}$, then $X$ has a non-empty boundary region $\overline{X} - \underline{X}$. This region contains borderline cases that $X$ might describe. I call such an $X$ a vague information set. Natural language works in this way: each word has a typical meaning referring to core states but can extend to borderline states in context, thereby lacking a sharp boundary.

## 3. Results

I assume the individual's expressed information set $X$ faithfully reflects her knowledge: $X$ includes all states she cannot distinguish from the core and excludes all states she can.[8] That is, $X$ corresponds exactly to the cluster of states that remain indistinguishable under her knowledge $U$. Under this assumption, I obtain two fundamental properties of vague knowledge and expression (proofs are in the Appendix).

**Proposition 1 (Informative Value of Vague Knowledge).**

$$\textit{If } \underline{X} \neq \emptyset \textit{ and } \overline{X} \neq \Omega, \textit{ then } \exists \omega_1 \in X \textit{ and } \omega_3 \notin X \textit{ such that } \omega_1 \nsim \omega_3.$$

Proposition 1 states that an information set is informative as long as it has some definite content and it does not trivially include everything. In other words, knowledge need not be precise to be informative. Even vague knowledge reduces uncertainty if it allows the individual to distinguish at least one state within $X$ from at least one state outside $X$. While the scope of uncertainty reduction may be limited, this still represents an informational gain relative to complete uncertainty.

[8] For strategic vague communication, see Lipman (2025).

**Proposition 2 (Expression of Vague Knowledge).**

*If* $\sim$ *is transitive on* $X$*, then* $\underline{X} = \overline{X}$*; If* $\sim$ *is non-transitive on* $X$*, then* $\underline{X} \subset \overline{X}$.

Proposition 2 states that precise knowledge can be expressed precisely, whereas vague knowledge cannot. Any faithful description of vague knowledge must have a non-empty borderline region. The intuition follows from dimensionality. Knowledge $\Omega \times \Omega \backslash U$ resides in a two-dimensional space of state pairs (it concerns which pairs are distinguishable), whereas an expressed information set $X$ is a one-dimensional subset of $\Omega$. Under precise knowledge, this projection loses no information because the equivalence classes serve as a sufficient statistic for the transitive relation. Under vague knowledge, however, the non-transitive relation cannot be fully captured by any single crisp subset. A vague expression, with its indeterminate boundary, partially accommodates that extra dimension in indistinguishability. Thus, vague knowledge is communicable through vague expression: natural language can convey aspects of knowledge that numbers cannot.

## 4. Conclusion

I present a simple model of information that incorporates vagueness. It formalizes vague knowledge (an inability to distinguish states with small differences) as a non-transitive indistinguishability relation. It also formalizes vague expression (an inability to represent a well-defined set of states) as a non-partitional information structure, characterized by the lack of sharp boundaries and the existence of borderline cases. My results imply that when the truth is only vaguely known, it is not amenable to precise definition or quantitative measurement, yet it remains communicable in natural language. In this sense, natural language conveys unmeasurable truths.

## Appendix: Proofs

### Proof of Proposition 1.

Consider any $\omega_1 \in \underline{X}$ and any $\omega_3 \notin \overline{X}$, Since $\underline{X} \subseteq X \subseteq \overline{X}$, $\omega_1 \in X$ and $\omega_3 \notin X$. Assume, for contradiction, that for a given $\omega_1$ I had $\omega_1 \sim \omega_3$. By the definition of $\overline{X}$, $\omega_3 \in \overline{X}$. This contradicts the choice of $\omega_3$. Therefore, $\omega_1 \nsim \omega_3$. ■

### Proof of Proposition 2.

The case of a transitive $\sim$:

Consider any $\omega_1 \in \overline{X}$. By the definition of $\overline{X}$, $\exists \omega_2 \in \underline{X}$ such that $\omega_1 \sim \omega_2$. By transitivity of $\omega_2$, $\forall \omega_2' \in \underline{X}, \omega_1 \sim \omega_2'$, satisfying the definition of $\underline{X}$. This implies $\omega_1 \in \underline{X}$. Therefore, $\underline{X} = \overline{X}$.

The case of a non-transitive $\sim$:

Case 1: $\underline{X}$ is a singleton.

Let $\underline{X} = \{\omega_1\}$. There exists $\omega_2' \neq \omega_1$ such that $\omega_1 \sim \omega_2'$. By the definitions of $\overline{X}$, $\omega_2' \in \overline{X}$. However, since $\underline{X}$ is a singleton containing only $\omega_1$ and $\omega_2' \neq \omega_1$, I have $\omega_2' \notin \underline{X}$. Therefore, $\underline{X} \subset \overline{X}$.

Case 2: $\underline{X}$ is not a singleton.

Consider any $\omega_1, \omega_2 \in \underline{X}$ and $\omega_1 \sim \omega_2'$. By the definitions of $\overline{X}$, $\omega_2' \in \overline{X}$. By non-transitivity of $\omega_1$, there must exist $\omega_2 \nsim \omega_2'$, so that $\omega_2' \notin \underline{X}$. Therefore, $\underline{X} \subset \overline{X}$. ■

## Figure 1. Precise and Vague Information

This figure illustrates simple examples of knowledge and expression in a three-state space $\{a, b, c\}$. A dotted line between states indicates they are indistinguishable (uncertainty), while no line indicates the states are distinguishable (knowledge). An illuminated area represents an information set, enabling the receiver to distinguish states in light from those in shadow. (A) Complete mutual indistinguishability yields no information. (B) Transitive elimination of indistinguishability: state $a$ is distinguishable from both $b$ and $c$, producing an information set with a sharp boundary that partitions the state space. (C) Non-transitive elimination of indistinguishability: $a$ is distinguishable from $c$ but not from $b$. Consequently, the information set containing $a$ must have a boundary region (penumbra) that admits the borderline state $b$. The resulting structure is a cover rather than a partition of the state space.

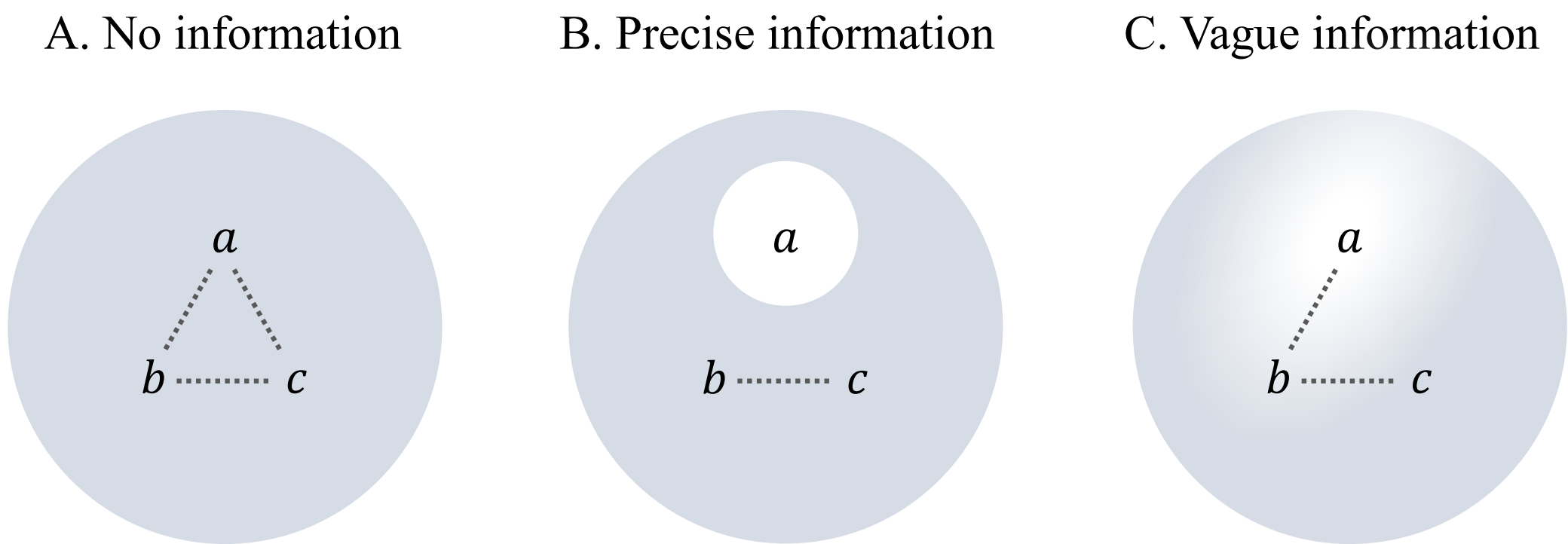